\documentclass[11pt]{article}
\usepackage{tabularx}
\usepackage{graphicx}
\graphicspath{{figure/}}
\usepackage{amsmath}
\usepackage{xcolor}
\usepackage{subcaption}
\usepackage[bookmarks=true]{hyperref}
\usepackage[margin=1.1in]{geometry}
\usepackage{float}
\usepackage{siunitx}

\newcommand{\basemdl}[1]{\texttt{SinoTx}}
\newcommand{\svtx}[1]{\texttt{SVTx}}
\newcommand{\ctx}[1]{\texttt{CTx}}
\newcommand{\dntx}[1]{\texttt{DnTx}}
\newcommand{\mwtx}[1]{\texttt{MWTx}}
\newcommand{\tomogan}[1]{\texttt{TomoGAN}}
\newcommand{\iradon}[1]{\texttt{iradon}}

\title{\bf \Large Masked Sinogram Model with Transformer for ill-Posed Computed Tomography Reconstruction: a Preliminary Study}

\date{\vspace{-5ex}}
\author{Zhengchun Liu, Rajkumar Kettimuthu, Ian Foster\\Data Science and Learning division, Argonne National Laboratory\\ zhengchun.liu@anl.gov}

\begin{document}
\maketitle
\begin{abstract}
Computed Tomography (CT) is an imaging technique where information about an object are collected at different angles (called projections or scans).
Then the cross-sectional image showing the internal structure of the slice is produced by solving an inverse problem. 
Limited by certain factors such as radiation dosage, projection angles, the produced images can be noisy or contain artifacts.
Inspired by the success of transformer for natural language processing, the core idea of this preliminary study is to consider a projection of tomography as a word token, and the whole scan of the cross-section (A.K.A. sinogram) as a sentence in the context of natural language processing.
Then we explore the idea of foundation model by training a masked sinogram model (MSM) and fine-tune MSM for various downstream applications including CT reconstruction under data collections restriction (e.g., photon-budget) and a data-driven solution to approximate solutions of the inverse problem for CT reconstruction. Models and data used in this study are available at \url{https://github.com/lzhengchun/TomoTx}.
\end{abstract}

\section{Introduction}
\subsection{Computed Tomography}
X-ray light source facilities generate massive data that give opportunities for AI/ML techniques to accelerate science discovery by processing X-ray data more efficiently and accurately.
However, the challenge is that there is no good way to make all the data publicly available to the community. 
We thus propose to train task-agnostic \textbf{foundation ML models} privately on data and only share the trained models to the AI/ML community as foundations for building ML solutions to process new X-ray data generated at light-source facilities. 

\begin{figure*}[ht]
\includegraphics[width=\textwidth]{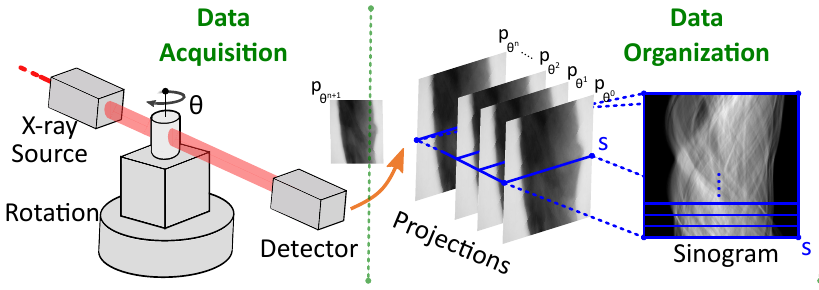}
\caption{Tomographic data acquisition and data organization. Figure credit: Bicer et al.~\cite{liu2019deep}.}
\label{fig:tomopipeline}
\end{figure*}

\autoref{fig:tomopipeline} illustrates the tomographic data acquisition, management, and analysis phases at synchrotron light sources.  
During the data acquisition phase, a sample is placed on a rotation stage and illuminated by X-ray. 
As X-rays pass through the sample, the photons---attenuated to a degree determined by the thickness and density of the object---are measured by the detector. 
The corresponding measurement is called a {\em projection}.  
A tomography experiment collects projections from different rotations ($\theta$), with typically a fixed exposure time for each.
An ideal experiment collects projections, $P=\{P_{\theta^0}, P_{\theta^1}, \dots, P_{\theta^n}\}$, that fully cover the sample. 
This paper uses term ``scan'' and ``projection'' interchangeably. 
All projections for a particular slice is called a {\em sinogram}.
Reconstruction from projections collected in an ideal scenarios is straightforward and out the scope of this study.
This paper instead focuses on solving the reconstruction problem when the projections are limited either by rotation angle coverage (e.g., sparse view, missing wedge) or exposure time (e.g., low dosage).

Beer's law shows the underlying mathematical model for the measurement process~\cite{beer1852bestimmun}: 
\begin{equation}
I_\theta(s) = I_0(s) \exp\left[-p_\theta(s)\right],
\end{equation}
where $I_0(s)$ is the incident x-ray illumination on the sample and $I_\theta(s)$ are the collected measurements at a number of $\theta$s, as a result of a tomographic scan. 
$p_\theta(s)$ represents a cross section of projections (shown in blue in the central section of \autoref{fig:tomopipeline}), known as a {\em sinogram}. 
For parallel beam geometry, measurements in a sinogram correspond to a cross section of the target sample. 
The tomographic reconstruction process aims to recover 2D cross section images of a sample from their corresponding sinograms.


\subsection{Deep neural network for X-ray imaging}
Machine learning without human-labeled data (i.e., unsupervised and self-supervised learning) is a longstanding challenge. 
Recently, we have seen great success in natural language processing (NLP), as transformer models like BERT, GPT-3, RoBERTa, and other variants have achieved top performance on a wide array of language tasks. 
Similar models has been successful in producing strong features for natural images (e.g., ImageNet) processing, such as Image GPT, Swin transformer~\cite{liu2021swin}, and MAE~\cite{he2021masked}.
However, there is a big gap for science images that are distinct from natural images which usually have three 8-bit channels~(Red, Green, Blue) and are widely available on the Internet.
Science images generated at light source facilities are a visual representation of X-ray that reveals the morphology (e.g., through intensity) of the materials being studied. 

There is a growing body of work on AI/ML methods (especially from computer vision) for processing X-ray data for tomography, serial crystallography, X-ray diffraction, and source diagnostics themselves. 
Most of these AI/ML based image processing tasks are based on supervised learning in which high quality labels are needed to train the DNN models. 
However, in most cases the labels are not easy (or even impossible) to get due to the uniqueness of each scientific problem.
Powered by the advancement of purpose-built AI accelerators, today's DNNs can easily over-fit one million images in the ImageNet dataset, thus demanding hundreds of millions of labelled images to train generalizable and versatile DNNs. 
This appetite for data has been successfully addressed in NLP by self-supervised pre-trained models such as BERT and GPT-3~\cite{he2021masked}.

A \textit{foundation model} is a model trained on broadly available data at scale, by using self-supervised learning without labels, and can be adapted (e.g., transferred and fine-tuned) to a wide range of downstream tasks~\cite{bommasani2021opportunities}.
In self-supervision, a pretext task is defined such that labels can be automatically calculated directly from the raw data instances~\cite{raghu2020survey}. 
This pretext task is defined without needing any labelling effort, but can be used to teach the network to extract task-agnostic representations from raw data. 
Thus, the foundation model is a multi-purpose (or even general-purpose) model that is task-agnostic when training and can serve multiple downstream tasks through corresponding adaptation with minimum labelled samples. 
There is a growing amount of work on building foundation models for computer vision, inspired by the success of Transformers for NLP.
However, most of those efforts, developed from research supported through commercial or academic pathways, are aimed at natural images and NLP. 
Although we call X-ray data scientific images, they are distinct from natural images that are composed of red, green, and blue.

The foundation model is a neural network trained on massive unlabeled data~(e.g., those data collected at X-ray light-source facilities), either of the same modality or a permutation of multiple modalities. 
As the computing demand is big and the training data is massive, or may not be able to move out of the facility due to ownership policies in this stage, the training of foundation models can be performed at computing facilities using supercomputers or purpose-built AI systems~\cite{liu2021bridging}. 
The training, updating and indexing of foundation models are needed only when data availability increased significantly, and more importantly their timescale are much more relaxed thus can make use of spare resource~\cite{liu2021bftrainer} for such purpose. 
For a given downstream task, one can query to determine the best model for use as a foundation to adapt from (e.g., transfer learning, fine-tune) with many fewer labeled samples.
Then, the adaptation stage can be performed anywhere (e.g., locally at a user/developer's institution), as both dataset and computing requirement are much smaller.

Moreover, similar as the notion of federated learning on data privacy, one can train foundation models using SSL locally within the X-ray light source facility, and only share the trained model with the community. 

\section{A Foundation model of Sinogram}
\subsection{Method}
\begin{figure}[htb]
\centering
\includegraphics[width=\textwidth]{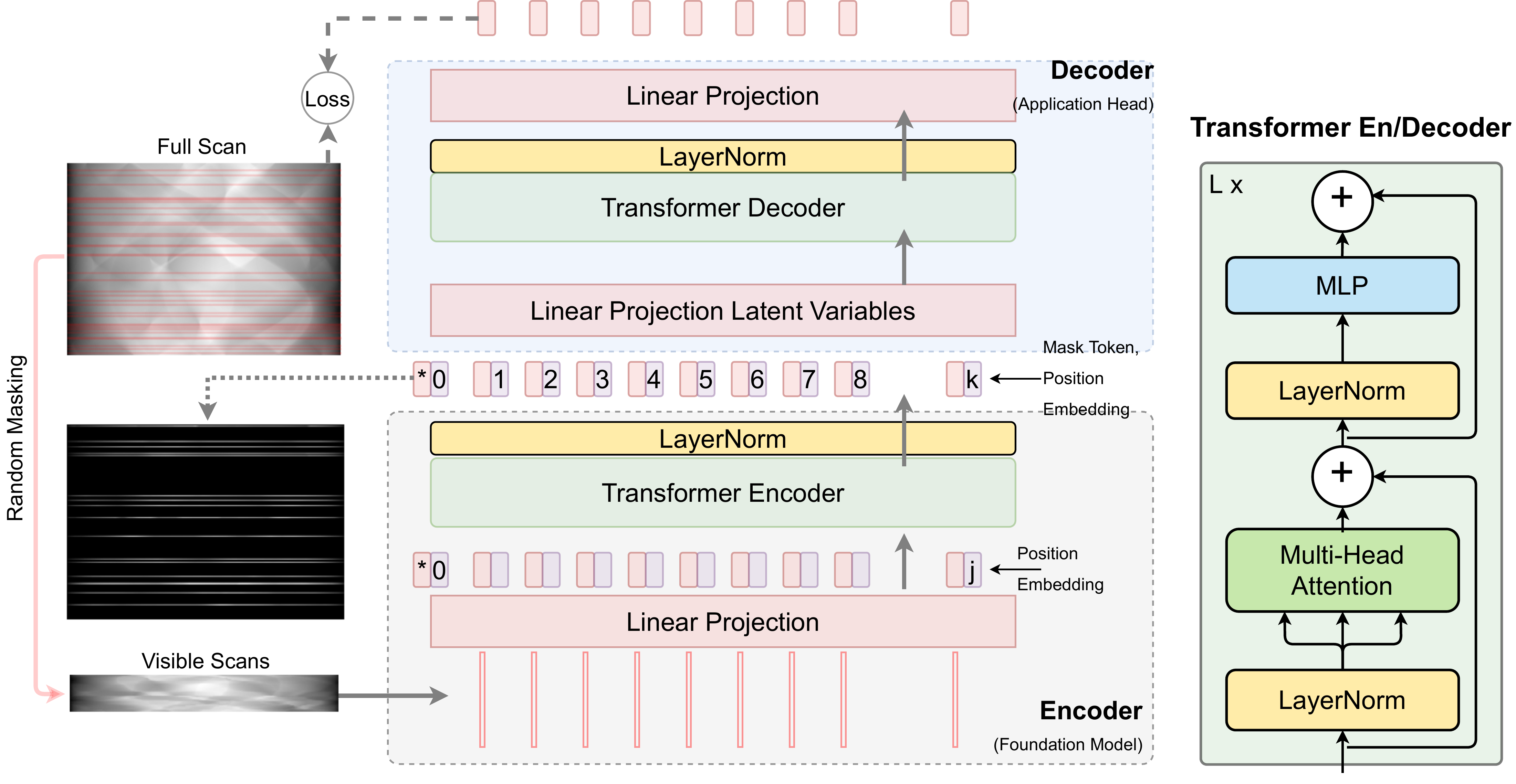}
\caption{Model overview of \basemdl{}. During pre-training, a large random subset of tomography views (e.g., 80\%) are masked out, i.e., on the left side of the figure only those projects with red overlay are kept. The encoder is applied to the small subset of selected views. Mask tokens are introduced after the encoder, and the full set of encoded views and mask tokens is processed by a small decoder to recover the original full tomography views in pixels (i.e., in-paint missed views). After pre-training, the encoder weights are frozen and applied to full tomography views and the decoder is replaced with fresh new model and trained for different downstream applications. The illustration of the Transformer encoder/decoder was inspired by Vaswani et al.~\cite{vaswani2017attention} and Dosovitskiy et al.~\cite{dosovitskiy2020vit}.}
\label{fig:big-pic}
\end{figure}

Similar as masked language model, we call this Masked Sinogram Model (MSM), once trained, it can be used as a foundation for down-streaming tasks. 
We open source our PyTorch based implementation \basemdl{} here \url{https://github.com/lzhengchun/TomoTx}.
The synthesized dataset that we used in this paper can be generated using script provided in our open source repository (\texttt{dataset/simu.py}).

\subsection{Results}
\begin{figure}[htb]
\centering
\includegraphics[width=0.85\textwidth]{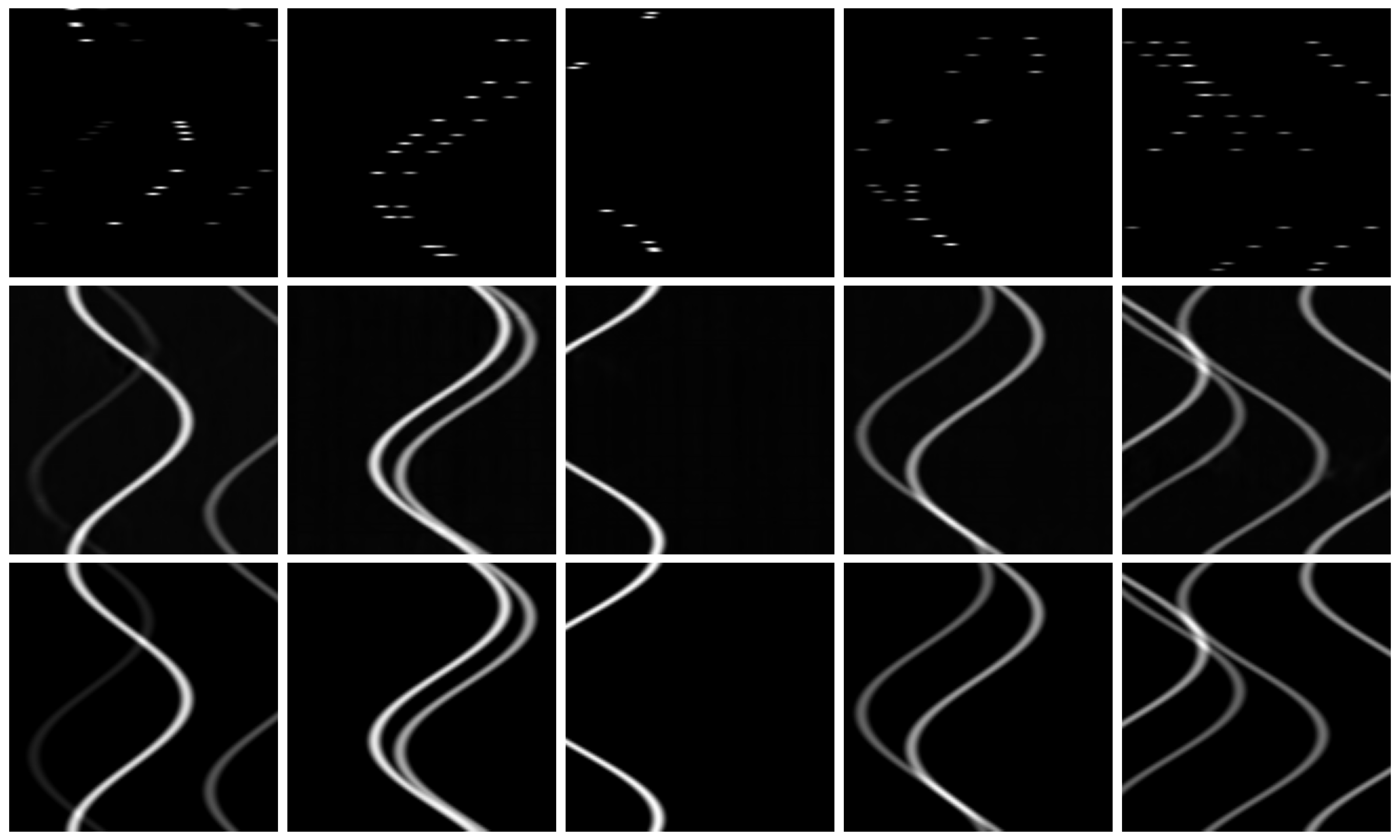}
\caption{R1: randomly masked sinogram; R2: in-painted sinogram; R3: Full-view sinogram. Note (the same for all figures of its kind in this paper): Each column is a randomly selected sample from the hold-out evaluation dataset, each row (called R1, R2, $\ldots$ thereafter) shows a specific representation of each sample. }
\label{fig:CB-cmp}
\end{figure}

The CookieBox detector~\cite{Audrey2019} is an angular array of electron Time-of-Flight (eToF) spectrometers. 
The x-ray shot photo-ionizes gas molecules in the interaction point, ejecting electrons. 
These electrons drift through a series of electrostatic potential plates and then detected by microchannel plates (MCP). 
So, instead of rotating samples to measure projections at different angle, here we have detectors placed at different angle so that projections are collected without rotation.
And the output will be an image containing the probability density of electrons' energy in each channel.
This problem becomes difficult when the number of detected electrons is low. 
Here we train the \basemdl{} by masking out measurements of some randomly selected MCP.
Once trained, e.g., using simulation data with ground truth label, we can use \basemdl{} to recover complete measurement from limited channels.
\autoref{fig:CB-cmp} presents randomly selected testing samples of a relatively simple scenario where the image represent histogram of energy from CookieBox detector. 
As one can see, \basemdl{} is able to recover the missing angle perfectly

\begin{figure}[H]
\centering
\subfloat[Sinogram. R1: randomly masked; R2: in-painted; R3: Full-view.]{\label{subfig:teaser-sino}\includegraphics[width=0.85\textwidth]{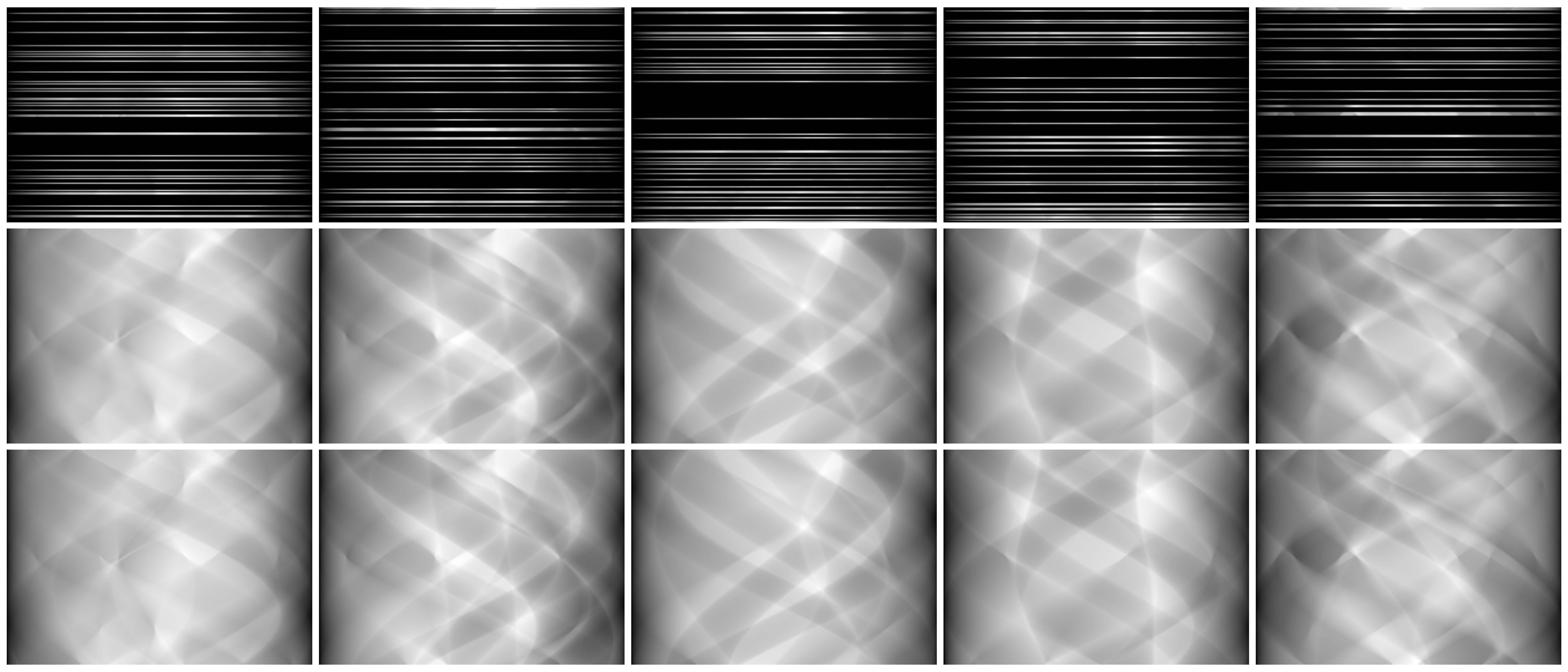}}\hfill
\subfloat[Reconstructions. R1: randomly masked view; R2: In-painted view; R3: Full-view.]{\label{subfig:teaser-recon}\includegraphics[width=0.85\textwidth]{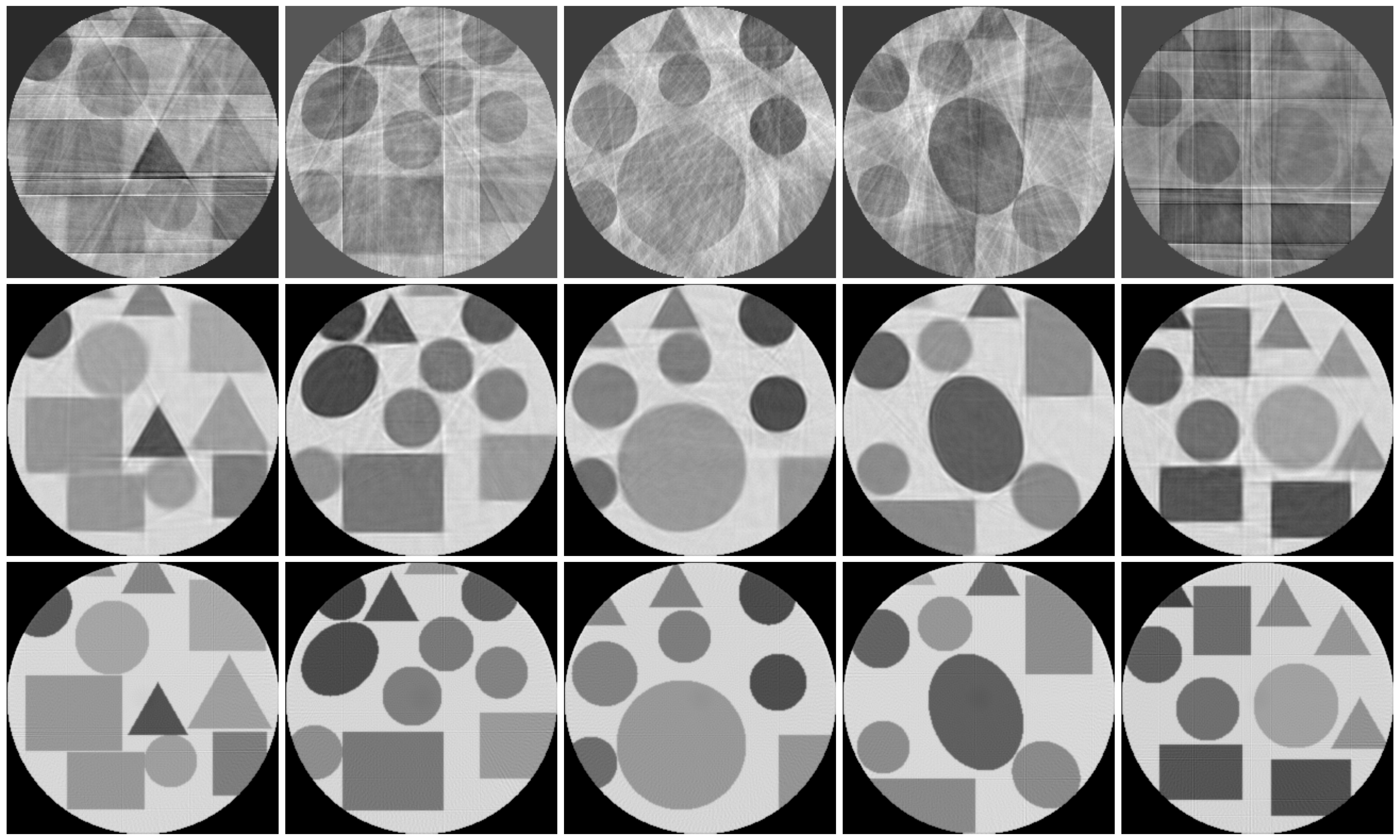}}
\caption{A comparison of full-view (180 evenly distributed projections from 0-180$^{\circ}$), randomly masked view (black rows are missed projections) and in-painted (using \basemdl{}) view computed tomography.}
\label{fig:teaser}
\end{figure}

As the final goal is the reconstruction, \autoref{fig:teaser} presents performance of \basemdl{} on a synthesized dataset with reconstructions. 
Specially, sinograms are simulated using Radon transform~\cite{Radon86} from synthesized images (i.e., mix of circle, ellipsoid, triangle and rectangle).
\autoref{subfig:teaser-sino} shows , visually there is no difference between in-painted sinograms and full view, however in-painting is not the final goal.
\autoref{subfig:teaser-recon} compares reconstructions, using inverse Radon transform (called \iradon{} thereafter), from different sinograms (i.e., direct masked views, in-painted views and full views).
As one can see, compare reconstruction from in-painting with reconstruction directly from randomly masked view, in-painted sinogram using \basemdl{} have much better quality (i.e., much less artifacts).
Moreover, although there is no noticeable difference between in-painting and full-view as shown in \autoref{subfig:teaser-sino}, their reconstructions do have visible difference. 
That being said, in-painting sinograms is not only a computer vision problem to focus on contrast only, instead it also needs to do well on pixel intensity (i.e., a regression problem).
Thus, although there is no need to include unmasked scans in the loss function calculation, we included the entire sinogram for loss (mean squared error) calculation so that the scale of the predicted sinogram will have consistent scale for further reconstruction.

\subsection{Model Explanation}
\begin{figure}[!ht]
\centering
\includegraphics[width=\textwidth]{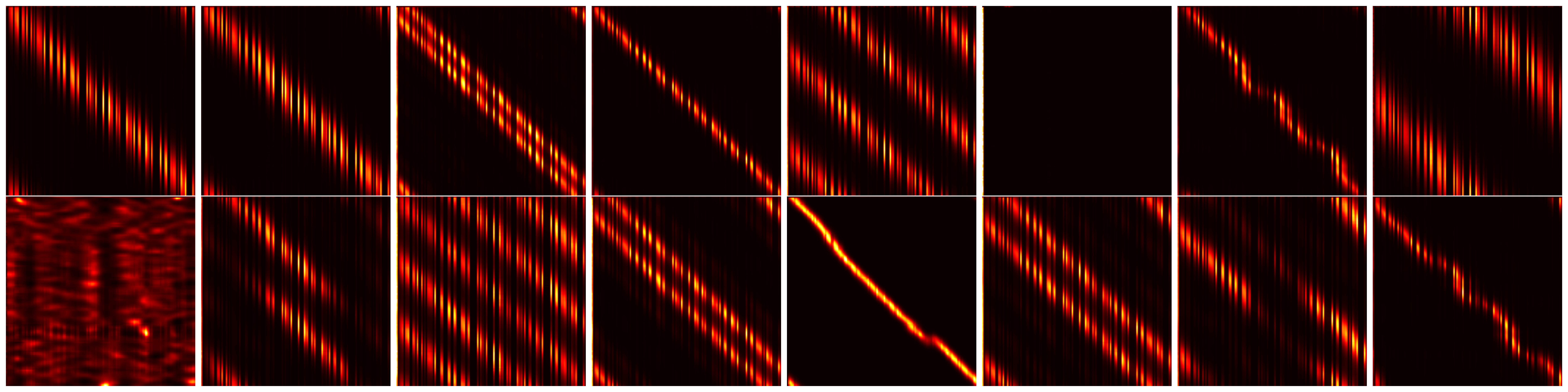}
\caption{Attention map of the 16 heads, of the first layer of the decoder.}
\label{fig:CB-atten}
\end{figure}

\autoref{fig:CB-atten} shows the attention map of the first decode layer where 50\% of the scans are randomly masked out as input. 
As one can see from \autoref{fig:CB-atten} that the dependence length of different heads are different, some only focus on neighboring sequence while others(e.g., the 9th and 11th head) focus on long distance dependence. 

\section{Downstream applications}
We explore three down-streaming applications on top of the masked sinogram model \basemdl{}: 
\begin{enumerate}
\item \svtx{} for sparse view of low dosage measurements;
\item \dntx{} for shorter exposure time of low dosage measurements; and 
\item \ctx{}, a pure data-driven method to computed tomography reconstruction. 
\end{enumerate}
The detail of the models and their training strategies can be found from our pyTorch based  open source implementation: \url{https://github.com/lzhengchun/TomoTx}.

We used a simulated dataset for the evaluation of four downstream applications and the comparison with the CNN based post-processing method.
Each synthesized image has a size of $256\times 256$ pixels and consists a mix of circle, ellipsoid, triangle and rectangle with different/random size and contrast/intensity. 
Tomography measurements of the images, i.e. sinograms, are simulated using Radon Transform~\cite{Radon86} with a step size of 1 degree for 180 degrees.
For comparison purpose, Inverse Radon Transform, called \iradon{} thereafter, is used as the counterpart for analytic reconstruction algorithm.
The simulation dataset contains \num{100000} samples for training and \num{2000} hold-out samples for evaluation. 
We further split the \num{100000} samples for training and validation (e.g., 90\% + 10\%) when there is a need for hyper-parameter optimization. 

\subsection{Sparse View}\label{sec:sparse-view}
Sparse view is a way to achieve low X-ray dose CT ~\cite{pelt2018improving,pelt2018mixed,liu2020tomogan,liu2019deep} by using normal exposure time for each projection but a larger step size for rotation angle $\theta$ as shown in \autoref{fig:tomopipeline}.
That is, all projections have high quality as of normal dosage, but the number of projects is much less. 
The only difference from \svtx{} and the base model \basemdl{} is that scans of input sinogram to \svtx{} is \textit{uniformly} masked (i.e., scanned using a larger step size) while input to the based model \basemdl{} are randomly masked (shown in \autoref{fig:big-pic}). 

\begin{figure}[htb]
\centering
\includegraphics[width=\textwidth]{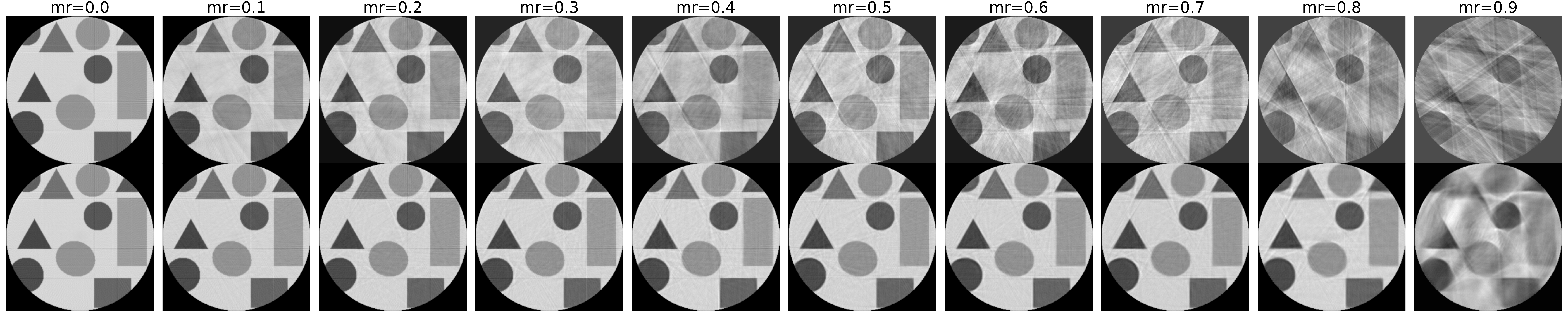}
\caption{Reconstruction from sparse sinogram with various masking ratio (upper row), and comparison with reconstruction from in-painted sinograms using \svtx{} (trained using mask ratio of 0.8).}
\label{fig:SinoTx-var-mr-cmp}
\end{figure}

We used exactly the same model architecture as \basemdl{} but only changed the way to mask projections of sinogram. 
Although \svtx{} can be a fine-tune of a trained \basemdl{}, here we retrain it from scratch with mean squared error loss using a fixed masking ratio of 0.8 (i.e., only keep 1 after every 4 projects). 
\autoref{fig:SinoTx-var-mr-cmp} compares reconstruction quality between directly using sparse-viewed sinogram and sinograms in-painted using \svtx{}.
As one can see, reconstructions from in-painted sinogram have much less artifacts. 
Moreover, \svtx{} trained with one pattern of sparse view (masked 0.8 angles) can generalize well to other less sparse patterns. 

\subsection{Short Exposure time Projection}
Another way, other than the sparse view (\S\ref{sec:sparse-view}), to achieve lower X-ray dosage is by using shorter exposure time for each projection thus keeps the same step size for $\theta$.
That is, shorter exposure time data collection scheme leads to the same number of projections as the normal dosage scheme but all projections have lower signal-to-noise ratio (SNR).
Conventional methods usually add regulation (e.g., total variation ) to the reconstruction algorithm to mitigate the low SNR, here we propose \dntx{} to denoise low-dose measurement in \textit{sinogram space}, or use CNN to denoise reconstructions as reconstructions usually have larger spatial-correlation than sinograms~\cite{liu2020tomogan,pelt2018improving,pelt2018mixed}, or at most to denoise projections~\cite{wu2020deep}.
To the best of our knowledge, this is the first work that directly work on denoising sinogram. 
The low dose sinograms used to train \dntx{} are simulated using Poisson distribution as used in \cite{wu2020deep} and \cite{leuschner2021lodopab}.
\begin{figure}[htb]
\centering
\includegraphics[width=\textwidth]{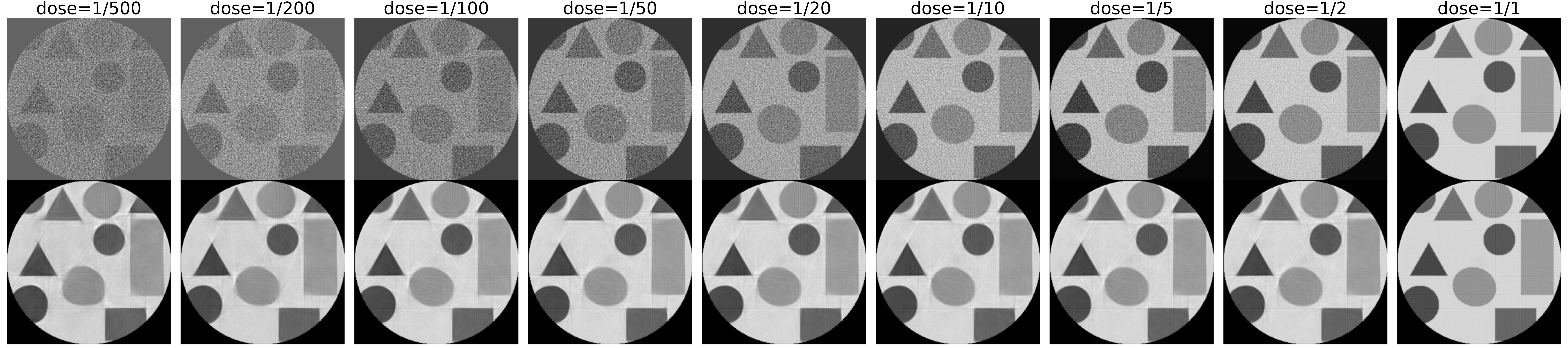}
\caption{Reconstruction from low dose sinograms (upper row), with various dosage relative to normal dose (right-most column), and comparison of reconstructions (bottom row) from corresponding sinograms denoised using \dntx{} (trained using 1/200 of dose).}
\label{fig:DnTx-var-dose-cmp}
\end{figure}

The model architecture of \dntx{} is identical to \basemdl{} except that there is no masking mechanism for \dntx{}.
We train the model using mean-squared-error on our simulated dataset synthesized to mimic using 0.5\% of the X-ray dosage. 
\autoref{fig:DnTx-var-dose-cmp} compares reconstructions from denoised sinograms using \dntx{} with different measurements X-ray dosage.
As one can see, the reconstructions from denoised sinograms(bottom row) have much better than its counterpart directly on noisy sinograms.
Furthermore, \dntx{} trained using 0.5\% of dosage generalize pretty well to other higher dosage measurements, and reasonably good to an even lower dosage (the left-most column, 0.2\%). 


\subsection{Data driven Direct reconstruction}
Different with existing work which use deep learning (e.g., CNN) to enhance quality (e.g., denoise, artifacts removal) of images reconstructed under constrained conditions~\cite{pelt2018improving,pelt2018mixed,liu2020tomogan} using analytical solution, here we present a pure data driven solution, \ctx{}, based on the pre-trained foundation model, \basemdl{}, to directly predict reconstruction based on sinogram with restrictions such as sparse view.

\begin{figure}[htb]
\centering
\includegraphics[width=\textwidth]{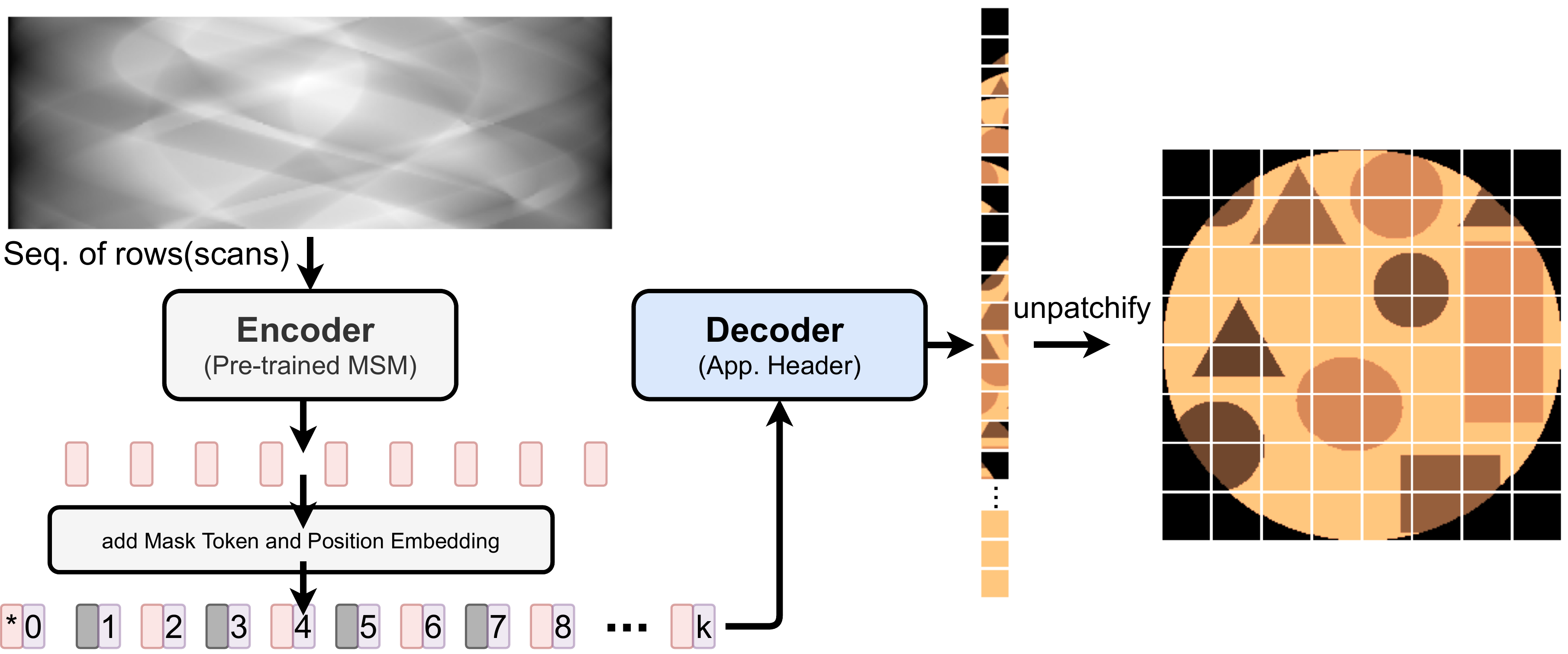}
\caption{Model overview of \ctx{}. The Encoder was pre-trained in the \basemdl{} via masked sinogram model (MSM) with a random mask ratio of 80\% (\autoref{fig:big-pic}). Mask tokens (train-able, illustrated as gray box) are introduced after the encoder, with \ctx{}'s Decoder (i.e., downstream application header), and the full set of encoded views (i.e., tokens) and mask tokens is processed by the decoder to reconstruct the image in real space. Note that, the Decoder predicts a sequence (default to an order of top-to-bottom and left-to-right) of patches of the image.}
\label{fig:ctx-arch}
\end{figure}

We kept the encoder of \basemdl{}, and only train the new decoder that is targeted to predict image in patches, similar as the decoder used in He et al.~\cite{he2021masked}.
\autoref{fig:ctx-arch} shows the architecture of the \ctx{}, different with other downstream applications discussed in this paper, \ctx{}'s Decoder predicts real images as a sequence of patches.
Particularly in this study with our synthesized dataset, the target image was split into $16\times16=256$ patches, each with $16\times16$ pixels.
So, the decoder will predict a sequence of 256 vectors, each has 256 items.  

\begin{figure}[htb]
\centering
\includegraphics[width=\textwidth]{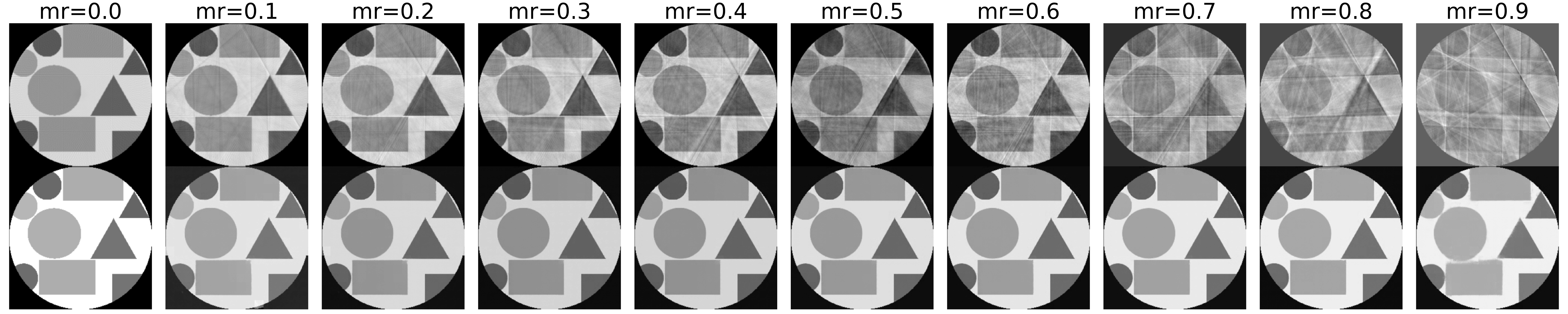}
\caption{Reconstruction directly from randomly masked sinogram, with various masking ratio, using \ctx{} (trained using mask ratio of 0.8).}
\label{fig:CTx-var-mr-cmp}
\end{figure}

\autoref{fig:CTx-var-mr-cmp} presents reconstructions using \ctx{} with different masking ratio to sinograms, and compared with reconstructions using analytical solution (inverse radon transform). 
As one can see, although \ctx{} is trained with a mask ratio of 0.8, it does perfectly well on smaller mask ratio and also make reasonable reconstruction with higher mask ratio. 
Moreover, it works much better than the analytical method especially when sinograms are very sparse.
It reveals that the model does learn the relation between sinogram and real-space image thus able to generalize to different missing wedges.

\begin{figure}[htb]
\centering
\subfloat[SSIM]{\label{subfig:CTx-vs-iradon-ssim}\includegraphics[width=0.5\textwidth]{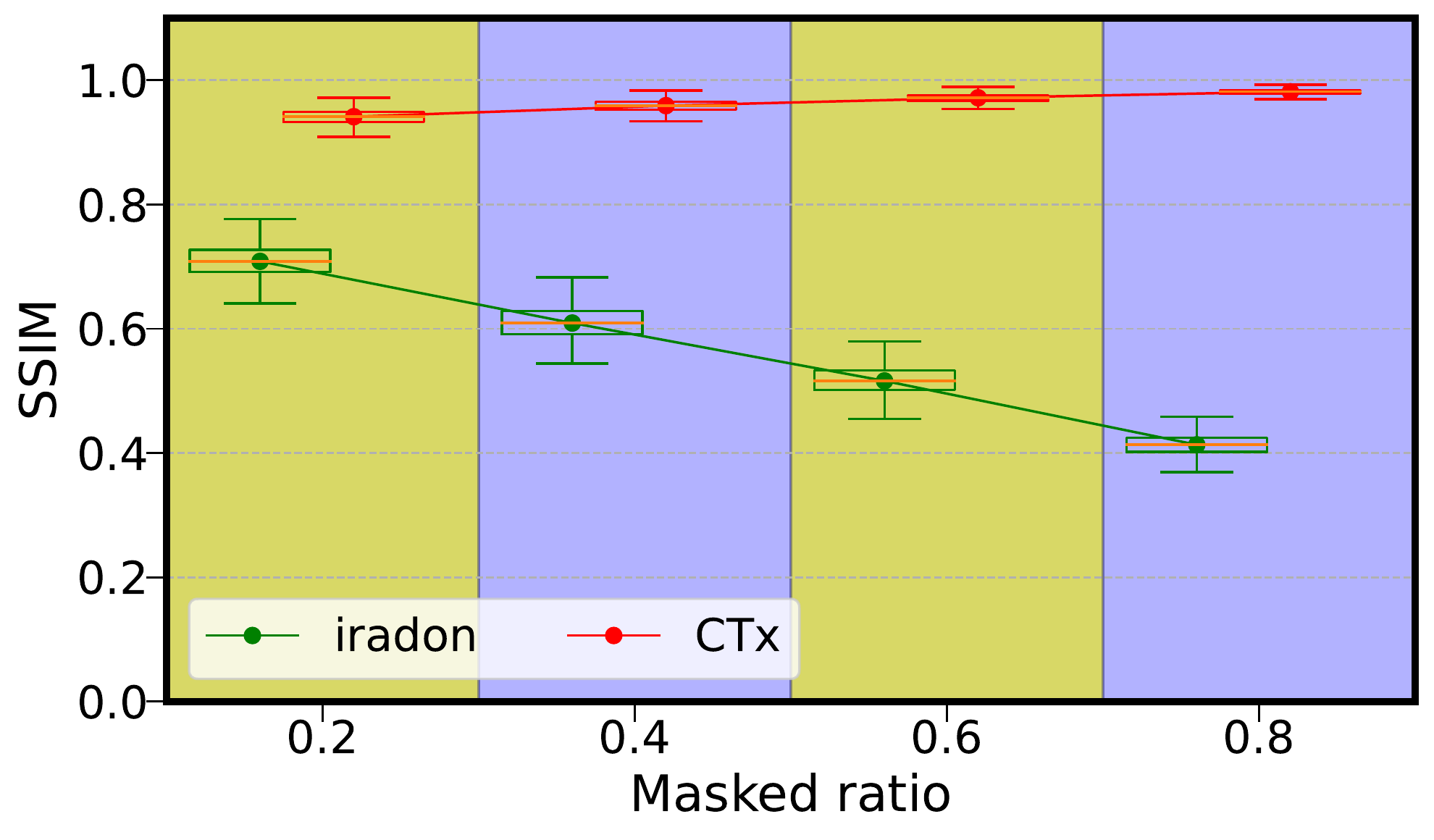}}
\subfloat[PSNR]{\label{subfig:CTx-vs-iradon-ssim}\includegraphics[width=0.5\textwidth]{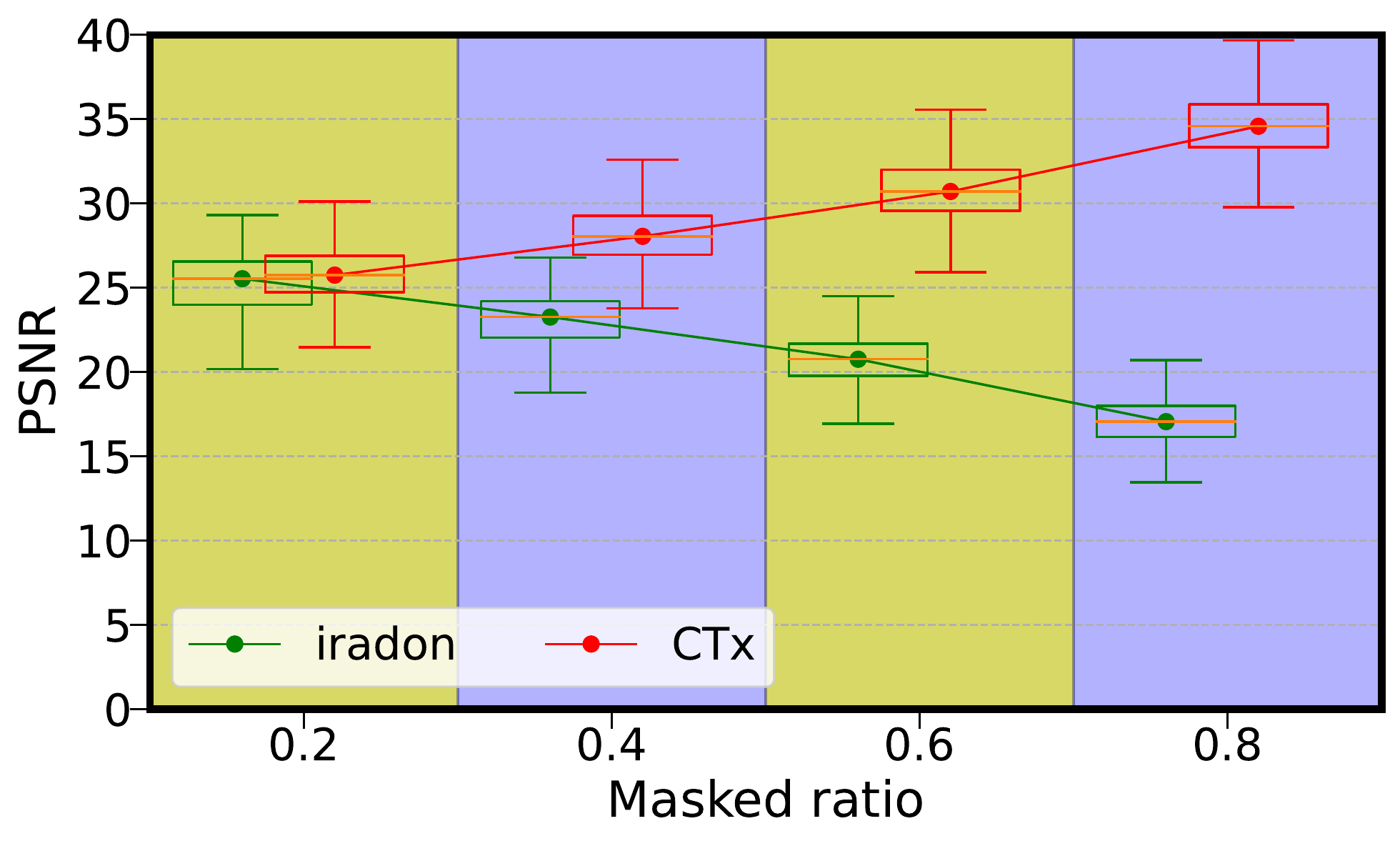}}
\caption{Quantitative evaluation of \ctx{} (trained with 0.8 masked ratio) using SSIM and PSNR.}
\label{fig:CTx-vs-iradon-ssim}
\end{figure}
\autoref{fig:CTx-vs-iradon-ssim} evaluates \ctx{} reconstruction quality quantitatively using image quality metrics: Structural Similarity Index(SSIM) and Peak signal-to-noise ratio (PSNR)~\cite{hore2010image}, based on \num{2000} samples (i.e., the evaluation dataset). 
As one can see, \ctx{} consistently performed better than \iradon{}  with different mask ratio. 
We note that, as \ctx{} was trained using 0.8 masked projections, the performance of \ctx{} dropped slightly to other mask ratio because of generalization error. 

\begin{figure}[htb]
\centering
\includegraphics[width=0.7\textwidth]{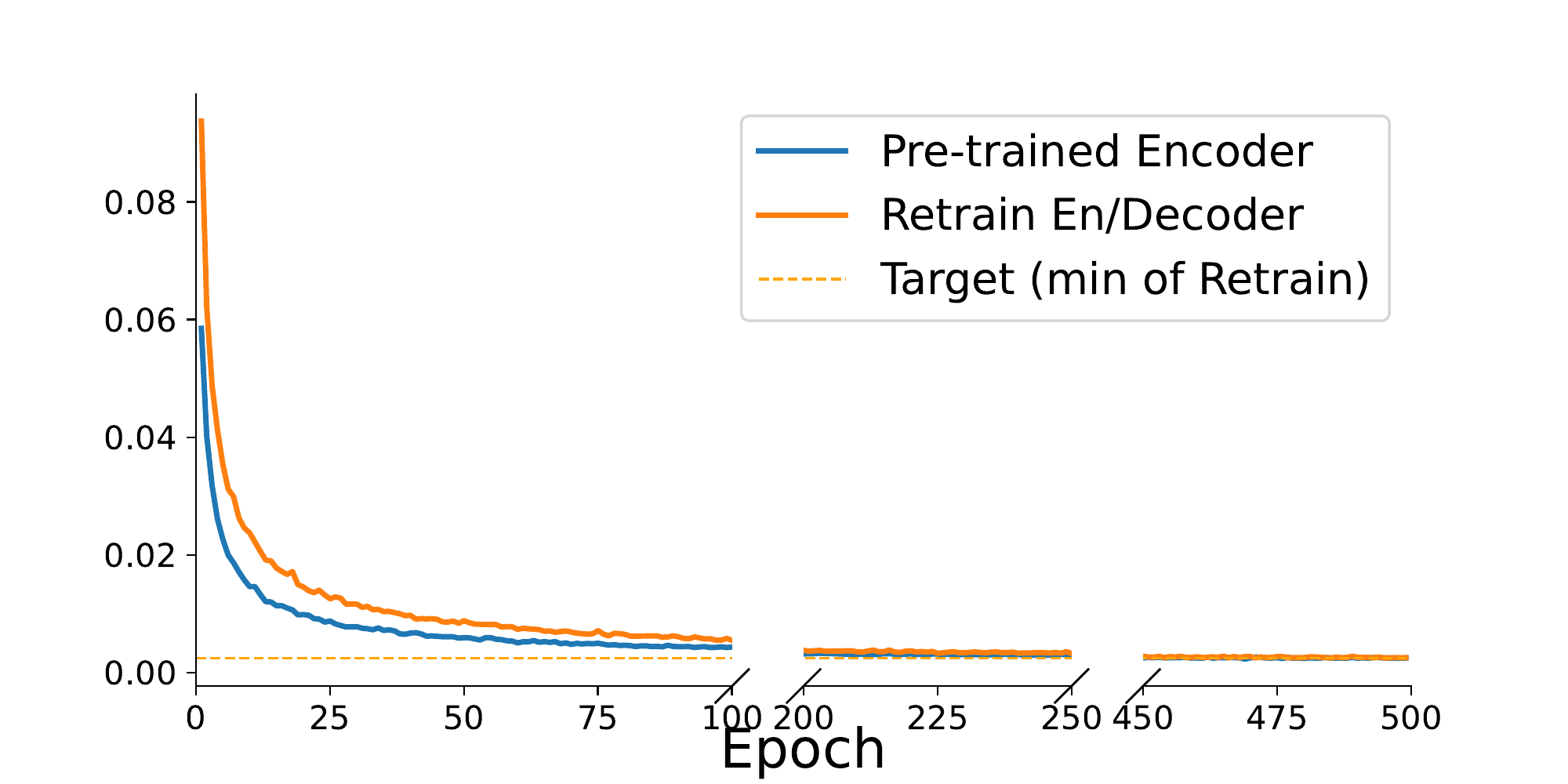}
\caption{A comparasion of model convergence curve between re-use pre-trained Encoder \& only train decoder, and retrain encoder \& new decoder from scratch. Y-axis is the validation loss.}
\label{fig:ft-compu}
\end{figure}

In terms of computing resource requirement and model convergence, \autoref{fig:ft-compu} compares the convergence rate between fine-tune (i.e., only train decoder) and retrain (for both encoder and decoder). 
As one can see, fine-tune converges much faster, it converges to a similar level as retrain using only about 20\% - 50\% of the epochs.

\section{Compare with CNN based post-process}
One of the most widely used methods to enhance tomography reconstruction with low-dose(LD) projects, sparse view(SV) or missing wedge(MW) measurements uses deep Convolutional Neural Network(CNN)~\cite{pelt2018improving,pelt2018mixed,liu2020tomogan}.
In which, CNN is trained in a supervised fashion using a pair of noisy reconstructions (e.g., reconstructed from LD, MW or SV) and ground-truth reconstruction (i.e., with normal measurements).
So, CNN works as a post-processing where conventional analytical methods (e.g., \iradon{}) are used for reconstruction and then get enhanced by the trained CNN. 
In this section, we will quantitatively compare results using proposed method in this study with state of the art CNN based solution. 

\begin{table}[htb]
\centering
\caption{Compare with CNN based post-processing solution (i.e., \iradon{}+\tomogan{}) in the case of sparse view with different (uniformly) mask ratio (MR) quantified using SSIM as a metric. All models are trained with MR=0.8.}
\begin{tabular}{c|ccccccccc}
\noalign{\hrule height 2pt}
Masked Ratio & 0.1 & 0.2 & 0.3 & 0.4 & 0.5 & 0.6 & 0.7 & 0.8 & 0.9 \\\hline
\iradon{} & \textbf{0.96} & 0.92 & 0.89 & 0.85 & 0.81 & 0.73 & 0.64 & 0.54 & 0.42 \\\hline
\tomogan{} & 0.93 & 0.93 & 0.93 & 0.93 & 0.92 & 0.92 & 0.92 & 0.92 & 0.82 \\\hline
\ctx{} & 0.93 & \textbf{0.95} & \textbf{0.96} & \textbf{0.96} & \textbf{0.97} & \textbf{0.98} & \textbf{0.98} & \textbf{0.99} & \textbf{0.95} \\\hline
\svtx{} & 0.88 & 0.83 & 0.82 & 0.81 & 0.81 & 0.80 & 0.75 & 0.90 & 0.33 \\\hline
\noalign{\hrule height 2pt}
\end{tabular}
\label{tbl:compare-with-cnn}
\end{table}

\tomogan{} solution used \tomogan{} to enhance (remove artifacts of) \iradon{} output. 
The analytical solution works the best when measurements are not so sparse (i.e., data collected in ideal condition) but dropped significantly when measurement becomes sparser. 
In comparison, NN based solutions, including state-of-the-art CNN, can maintain reconstruction quality well when increasing sparsity of measurement. 
It is important to note that, \ctx{}, our direct data-driven solution, performed consistently the best in all sparse view cases, exception to mask ratio of 0.1 which may not be considered as sparse.

\section{Conclusion and Future work}
Inspired by the success of natural language processing (NLP) using neural network, this preliminary work explored the idea to treat computed tomography as an NLP problem using a transformer based masked sinogram model (MSM).
We have seen promising results on the fine-tune of foundation MSM for different downstream tasks, including directly approximate a solution of the inverse problem by fine-tune a pre-trained MSM which to the best of our knowledge is the first pure data-driven solution for CT reconstruction. 

However, the dataset used in this study is relatively simple compare with real dataset, we envision (e.g., based on findings of Kaplan et al.~\cite{kaplan2020scaling}) that a much bigger model, more data and computing are needed to process real experimental dataset. 

\section*{Acknowledgements}
This material was based upon work supported by the U.S. Department of Energy, Office of Science, under contract DE-AC02-06CH11357.
This research used resources of the Argonne Leadership Computing Facility, a DOE Office of Science User Facility supported under Contract DE-AC02-06CH11357.
We thank Tekin Bicer and Ryan Coffee for the discussion of some of the results presented in this paper. 

\bibliographystyle{abbrv}
\bibliography{refs}

\section*{Government License}
The submitted manuscript has been created by UChicago Argonne, LLC, Operator of Argonne National Laboratory (``Argonne''). Argonne, a U.S.\ Department of Energy Office of Science laboratory, is operated under Contract No.\ DE-AC02-06CH11357. The U.S.\ Government retains for itself, and others acting on its behalf, a paid-up nonexclusive, irrevocable worldwide license in said article to reproduce, prepare derivative works, distribute copies to the public, and perform publicly and display publicly, by or on behalf of the Government.  The Department of Energy will provide public access to these results of federally sponsored research in accordance with the DOE Public Access Plan. http://energy.gov/downloads/doe-public-access-plan.

\end{document}